\documentclass{iau_FM}
\usepackage{graphicx,natbib}

\title[FM 17.~How to separate the low amplitude $\delta$~Scuti variation] 
{How to separate the low amplitudeδ\ $\delta$~Scuti variation from the 
instrumental ones in CoRoT data?}

\author[J\'ozsef M. Benk\H{o} \& Margit Papar\'o]   
{J\'ozsef M. Benk\H{o}
 \and Margit Papar\'o}

\affiliation{$^1$
Konkoly Observatory, MTA CSFK, Konkoly Thege M. u. 15-17., H-1121 Budapest, Hungary
\\email: {\tt benko@konkoly.hu}}

\pubyear{2015}
\jname{Astronomy in Focus, Volume 1} 
\editors{Piero Benvenuti, ed.}
\begin{document}

\maketitle

\begin{abstract}
Rich regular frequency patterns were found in the Fourier spectra of low-amplitude 
$\delta$~Scuti stars observed by CoRoT satellite (see Papar\'o et al. 2016a,b). 
The CoRoT observations are, however, influenced 
by the disturbing effect of the South Atlantic Anomaly. The effect
is marginal for high amplitude variable stars but it could be critical in the case of low
amplitude variables, especially if the frequency range of the intrinsic variation overlaps the
interval of the instrumental frequencies. Some tests were carried out both on synthetic and
real data for distinguishing technical and stars' frequencies.
\keywords{methods: data analysis, stars: variables: delta Scuti, techniques: photometric}
\end{abstract}

\firstsection 
\section{Introduction}

The photometic space telescope CoRoT \citep{Baglin} moves 
on a polar low Earth orbit with an orbital period of 1 hour 43 minutes. 
The orbital plane is fixed with respect to the center 
of the Earth so (due to the rotation of the Earth) the satellite flies
for each circulation above different geographical lenghts.
This orbital property causes that the satellite passes regularly
through the South Atlantic Anomaly (SAA). The SAA is an area where the 
Earth's inner Van Allen zone comes closest to the surface.
It leads to an increased flux of energetic particles what makes
the CCD images to be noisy. The CoRoT satellite observated continousely
even in the SAA as well, but the data pipeline signed these data points 
by a non-zero flag. If we use data points with zero flag only
we get of the noisy data but we lose the continuous sampling which causes 
alias frequencies in the Fourier spectra. The alias structure needs 
careful study before we search for regular spacing in the frequency
lists of CoRoT $\delta$~Scuti stars.

\begin{figure}
\begin{center}
 \includegraphics[width=12.5cm]{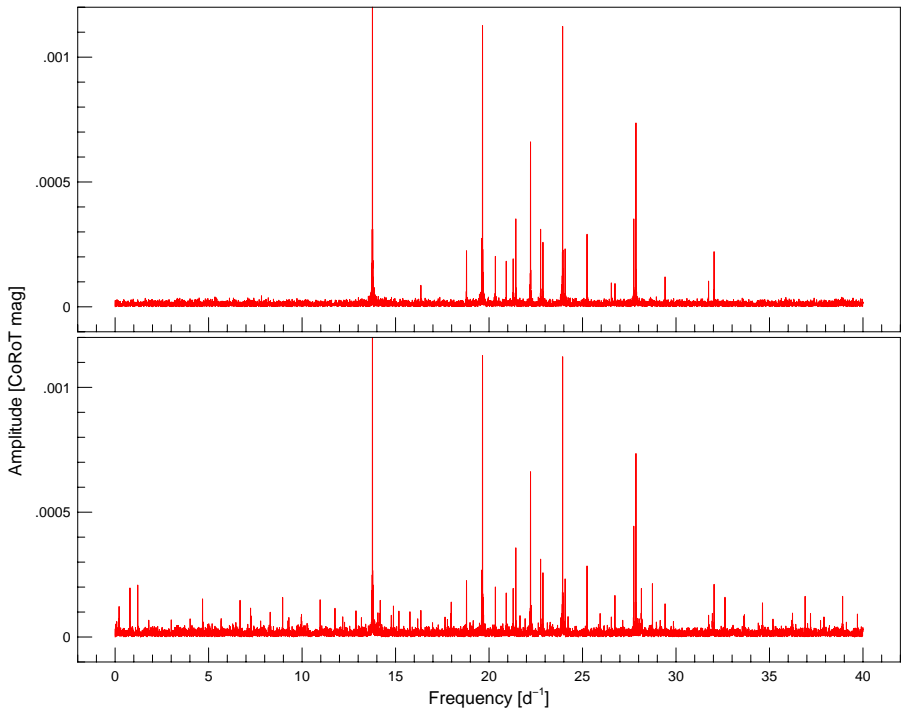} 
 \caption{Fourier spectra of artificial data. (top) The 22 input frequencies are seen in
the uniformly sampled data. (bottom) Many alias frequencies appear when we
simulate the real data: CoRoT sampling and removing the data points when the
satellite passing through the SAA.}
   \label{fig1}
\end{center}
\end{figure}

\section{Tests}

All statement here respect to EXO field stars in the 2nd CoRoT long run field located 
in the Galactic anticenter direction (LRa2) (see for the specifications \citealt{Auvergne}). 
We prepared an artificial light curve from 22 frequencies taken from the highest amplitude 
frequencies of a CoRoT $\delta$ Scuti star (CoRoT 102637079). 
The amplitudes were set to be identical with the observed ones
and Gaussian noise ($\sigma=1\times 10^{-5}$~mag) was also added to the sythetic light curve.
(This noise level is typical for a 11-13~mag star in the EXO field.)
The light curve was sampled with the CoRoT sampling time (1) either uniformly 
or (2) in those data points where CoRoT pipeline yields 0 flag (viz. without any technical problems). 
The Fourier spectra of these two synthetic time series are given in Fig.\,\ref{fig1}.

\begin{figure}
\begin{center}
 \includegraphics[width=\textwidth]{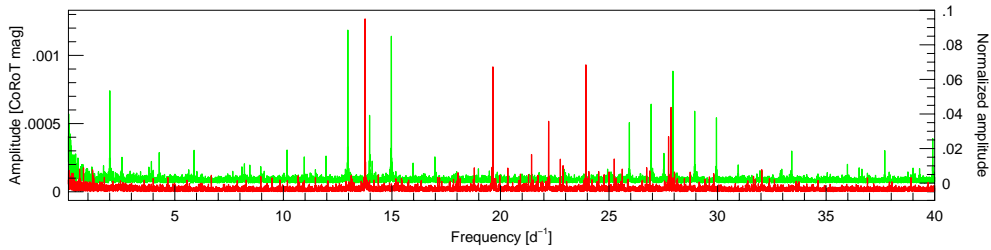} 
 \caption{The Fourier spectrum of the $\delta$~Scuti star CoRoT 102637079 (red). 
As a comparison we show the window function of these data (green). 
The two functions are plotted in different
scales: Fourier amplitudes are in CoRoT flux (left scale), while the window function 
is represented in the usual normalized scale (right scale).
}
   \label{fig2}
\end{center}
\end{figure}

The equidistantly sampled data has clear spectrum, that is no other peaks then the originally
set 22 ones (top panel in Fig.\,\ref{fig1}). 
While the spectrum of flag 0 data shows many low amplitude alias peaks caused 
by the non-uniform data sampling
(bottom panel in Fig.\,\ref{fig1}).
The alias phenomenon is well-known for ground-based data where daily and yearly (seasonal) alias
frequencies are the most frequent.  
The spectral window function of flag 0 point of the CoRoT data 
(green spectrum in Fig.\,\ref{fig2}) contains frequencies at $kf_{\mathrm{s}}+lf_{\mathrm{o}}$, where 
$f_{\mathrm{s}}=1.0027$~d$^{-1}$ (sidereal-day
frequency), $f_{\mathrm{o}}=13.97$~d$^{-1}$(orbital frequency), and $k, l$ are integers.
The amplitude of a given alias frequency is varied form star to star but Fig.\,\ref{fig2}
shows a typical situation: the highest peak amongst 
the sideral-day harmonic frequencies is the 2$f_{\mathrm{s}}=2.0054$~d$^{-1}$.
The largest order detectable harmonic is generally the 4th or the 5th one.
The orbital frequency amplitude is lower than  the amplitude of its first harmonic.
The amplitudes of the linear combination frequencies $f_{\mathrm{o}}\pm f_{\mathrm{s}}$
are higher than the amplitude of the orbital frequency $A(f_{\mathrm{o}})$.

The common pre-whitening processes produce further technical frequencies and increasing noise
level \citep{Balona2014b}, however, due to the many above mentioned aliases, 
a pre-whitening method is essential for handling these data. 
We used SigSpec program \citep{Reegen2007} for consecutive pre-whitening process.
The SigSpec algorithm with its default spectral significance limit ($sig=5$)
resulted in the 22 input frequencies for both artificial data sets.
Both the residual spectra indicate no significant peaks above the 0.01~mmag 
average (noise) level after the pre-whitening process. 
These results are consonant with \cite{Balona2014b} diagnose, namely that 
the higher amplitude frequencies are less affected by the faults of pre-whitening process.
In the same time we found that all of the used input frequencies can be considered as 
`higher amplitude frequency' though they amplitudes are between 1.5 and 0.1~mmag.

The red line in Fig.\,\ref{fig2} shows the spectrum of CoRoT~102637079.
This spectrum is very similar to the synthetic one (in the bottom panel of Fig.\,\ref{fig1})
which is evident because it was generated on the basis of the highest amplitude frequencies
of the real star. More instructive is how heavily overlaps the real star's frequency range with 
the location of the peaks of the spectral window. Despite the previous positive test result
this warns for caution. The observed data have additional biases from the sythetic case. 
The flag 0 data might be still encumbered by the influence of the SAA. The strength of this influence 
 differs from star to star. According to our experiences the alias structure can not be 
removed totally by a consecutive pre-whitening process.
The probable reason is that the proximity of the SAA causes some flux variations 
(an increase before the SAA and a decrease after that) 
which have different time lengths and amplitudes for each passing through.

Our conclusion is that the higher amplitude peaks of the Fourier spectra
can be easily separated from the instrumental ones. 
Attention has to pay to the increasing noise level is due to many-steps pre-whitening 
processes \citep{Balona2014b}. Any well-established study can be accepted only for the higher amplitude peaks. 
They should be referred as intrinsic frequencies of the star. Although, beacuse of the increased 
noise level, we may miss some of the low amplitude intrinsic frequencies.
When we found regular pattern using the high amplitude frequencies we may extend 
the pattern with lower amplitude frequencies.
This basic idea  -- using only the highest amplitude peaks -- was applied in 
 \cite{Paparo13} that in our new investigation on a larger sample of 
CoRoT $\delta$~Scuti stars we extended the regular patterns with lower amplitude frequencies, too.
The results and the details of the analysis will be published soon elsewhere (Papar\'o et al. 2016a, b).  

\acknowledgement{
This work was partially supported by the ESA PECS Grant No~4000103541/11/NL/KM.}


\begin{thebibliography}{}

\bibitem[Auvergne \etal\ (2009)]{Auvergne}
{Auvergne, M., Bodin, P., Boisnard, L. et al.} 2009,
\textit{A\&A}, 506, 411

\bibitem[Baglin \etal\ (2006)]{Baglin}
{Baglin, A.,  Auvergne, M., Boisnard, L., et al.} 2006, in
COSPAR Meeting, Vol. 36, 36th COSPAR Scientific Assembly, 3749


\bibitem[Balona (2014)]{Balona2014b}
{Balona, L.} 2014,
\textit{MNRAS}, 439, 3453

\bibitem[Papar\'o \etal\ (2013)]{Paparo13}
{Papar\'o, M., Bogn\'ar, Zs., Benk\H{o}, J.M., Gandolfi, D., Moya, A., et al.} 2013,
\textit{A\&A}, 557, A27 

\bibitem[Papar\'o \etal\ (2016a)]{Paparo16a}
{Papar\'o, M., Benk\H{o}, J.M., Hareter, M., Guzik, J.A.}, 2016a, \textit{ApJ} (accepted), astro-ph:1603.02050

\bibitem[Papar\'o \etal\ (2016b)]{Paparo16b}
{Papar\'o, M., Benk\H{o}, J.M., Hareter, M., Guzik, J.A.}, 2016b, \textit{ApJS} (accepted), astro-ph:1603.09161

\bibitem[Reegen (2007)]{Reegen2007}
{Reegen, P.} 2007,
\textit{A\&A}, 467, 1353

\end{thebibliography}
\end{document}